\documentstyle[12pt,aasms4]{article}
\begin{document}

\title{An Investigation of Be/X-ray Pulsars with OGLE-III Data } 

\author{ P.C.\ Schmidtke \& A.P.\ Cowley }

\affil{Dept.\ of Physics \& Astronomy, Arizona State Univ., Tempe, AZ,
85287-1504, USA; paul.schmidtke@asu.edu, anne.cowley@asu.edu } 

\and

\author{A. Udalski}

\affil{Warsaw University Observatory, Al.\ Ujazdowskie 4, 00-478 Warsaw, 
Poland; udalski@astrouw.edu.pl}

\begin{abstract}

We have studied five seasons of OGLE-III data for eight SMC Be/X-ray
pulsars for which no other survey data were available.  We have determined
orbital periods for four of these binary systems, one of which also shows
nonradial pulsations.  Optical identification of SMC X-2 is reconsidered,
but no periods were found for either of the two possible candidates. 

\end{abstract}

                                    
\section{Introduction}

Be/X-ray binary systems account for about half of the high-mass X-ray
binaries, but relatively little is known about their properties.  We have
been studying the photometric behavior of these systems in the Small
Magellanic Cloud using the OGLE-II and MACHO databases (Szymanski 2005,
Udalski et al.\ 1997, Zebrun et al.\ 2001, Alcock et al.\ 1999) in an
attempt to learn more about the component stars.  Many of these systems
contain X-ray pulsars; lists of such systems are given by Coe et al.\
(2005), Liu et al.\ (2005), Haberl \& Pietsch (2004), Yokogawa et al.\
2003,  and others.  Some of these binaries lie outside the fields of the
OGLE-II and MACHO surveys and hence had no available longterm photometry. 

The OGLE-III project covers a larger area around the SMC and includes some
Be/X-ray binaries for which longterm photometry had not previously been
obtained.  We have undertaken a study of some of these systems, as listed
in Table 1.  We give the OGLE-III positions of the optical counterparts
that are more accurate, in most cases, than previously available
positions.  Table 1 also lists the catalogue number in the list of
H$\alpha$ emission-line stars in the SMC by Meyssonnier \& Azzopardi
(1993; denoted as [MA93]) and the SXP number (Coe et al. 2005) which
gives the X-ray pulse period for \underbar{S}mall Magellanic Cloud
\underbar{X}-ray \underbar{P}ulsars.  Finding charts for all systems
studied here can be found on M.\ Coe's web
site\footnote{http://www.astro.soton.ac.uk/$\sim$mjc/smc}. However,
because of confusion about the optical identification of SMC X-2, we
present a new finding chart in this paper. 

\section{Analysis of Data from OGLE-III } 

The OGLE-III data cover five years with photometry in the $I$ bandpass.
Fig.\ 1 displays the longterm $I$ light curves for the sources studied
here.  For all systems the data were prewhitened (flattened) using a
low-order polynomial fit to remove long term variations in brightness.  In
some cases it was necessary to flatten the data in segments when there
were large, irregular brightness changes (e.g. SMC X-2).  We use $I^*$ to
designate the flattened data.  To analyze the $I^*$ data we computed
periodograms using the technique described by Horne \& Baliunas (1986). 
When the variations are very non-sinusoidal we also used the phase
dispersion minimization (PDM) technique of Stellingwerf (1978).  We
searched for photometric periods in the range of 0.25 days to 1000 days. 
The techniques we used here are more fully described in our earlier papers
on Be/X-ray pulsar binaries (e.g. Schmidtke et al.\ 2004, Schmidtke \&
Cowley 2005, Schmidtke \& Cowley 2006). 

\section{Individual Be/X-ray Pulsar Systems Studied with OGLE-III Data}

\subsection{ RX~J0052.9$-$7158 = SXP169 }

This pulsar is identified with the Be star [MA93]623.  Galache et al.\
(2005) note that SXP169 and SXP165 are the same pulsar which has been spun
up by $\sim$3 sec over 7.5 years. 

The OGLE-III light curve for SXP169 (Fig.\ 1) shows a small longterm
variation with considerable scatter about the mean.  Analysis of all
seasons of photometry shows two clear periods.  The first is the orbital
period of P=67.6$\pm$0.3 days, found both in the periodogram and in the
PDM variance (Fig.\ 2).  The epoch of maximum is JD~2452583.3.  The light
curve folded on this ephemeris shows an outburst of $\sim$0.02 mag with a
rapid rise and more gradual decline.  Galache et al.\ (2005) derived a
period of 68.6$\pm$0.2 days from their weekly X-ray monitoring program
with $RXTE$.  We note that both our period and outburst epoch are in
reasonable agreement with those of Galache et al. 

In addition to orbital outbursts, we also have found nonradial pulsations
(NRP) of the Be star with a period of P$_{NRP}$=0.755 days.  These
pulsations are best detected in the out-of-outburst data (i.e. from
orbital phases $\sim$0.2-0.9.  In Fig.\ 2 we plot the periodogram and
folded light curve for these NRP.  There is no evidence that the pulsation
period changed through the five seasons of OGLE-III observations. 

\subsection{ SMC X-2 = RX~J0054.5$-$7340 = SXP2.37 }

SMC X-2 is one of the earliest X-ray sources found in the SMC (Li,
Jernigan, \& Clark 1977), and it was soon noticed to be highly variable
(Li \& Clark 1977).  Sanduleak \& Philip (1977) identified SMC X-2 with a
14th magnitude OB star, and this was supported by spectroscopic
observations by Allen (1977) who found H emission lines.  Subsequent
spectra taken by Crampton et al.\ (1978) showed the star to be a late Oe
star with a velocity appropriate to the SMC.  However, Murdin, Morton, \&
Thomas (1979) later obtained an improved image of the field and resolved
the optical counterpart into a close double with $\sim$0.8 mag difference
in brightness.  They assigned spectral types of O to the brighter
(northern) star and Be to the fainter (southern) star.  Because the
components are very close together ($\sim2.5^{\prime\prime}$), it is
possible that the emission observed by both Allen and by Crampton et al.\
came from the fainter component which would have been blended with the
spectrum of the brighter star.  However, Oe and Be stars have long been
known to lose their emission lines from time to time, so we cannot rule
out the brighter component as being the X-ray source.  The X-ray position
is not sufficiently accurate to distinquish between these two close stars.
Because the only finding chart in the literature is a low resolution image
(Clark et al. 1978), we present the OGLE-III image in Fig.\ 3 which
clearly shows the two components.  In addition, we give the OGLE-III
position of each component in Table 1. 

Fig.\ 1 shows the light curves, using comparable scales, for both the
northern (brighter) and southern (fainter) component.  The northern star
has a swooping light curve with a range of $\sim$1 mag.  This type of
variation is commonly seen among the Be/X-ray pulsar binaries.  The
southern component is 1-2 magnitudes fainter and shows only a small light
variation.  Because the two components are so close, it is likely that the
``variability" of the southern star is due to contamination with the much
brighter northern component whose magnitude is extremely variable. 

For completeness, we have searched for periodic behavior in both
components.  The brighter, northern component shows weak periodicities at
$\sim$90, 46, and 23 days.  All are probably aliases of each other.  The
folded light curves show very small sinusoidal variation.  None is
convincing enough to claim as a true orbital period.  No periods were
found in the analysis of the fainter, southern star.  Based on the
character of the light curves, we conclude the northern star is more
likely to be the optical counterpart of SMC X-2, but an improved X-ray
position is needed to confirm this. 

\subsection{ 2E~0053.7$-$7227 = SXP34.1 }

This source was found to be a 34-sec X-ray pulsar by Edge et al.\ (2004)
using $Chandra$ data (CXOU~J005527.9$-$721058).  The excellent X-ray
position makes the optical identification secure.  As can be seen in Fig.\
1, the longterm light curve shows a relatively constant mean magnitude of
$I$=16.82 with a scatter of $\sim$0.06 mag.  We searched the entire data
set and each of the individual seasons for possible photometric periods,
but found nothing convincing. 

\subsection{ RX~J0057.8$-$7202 = SXP280 }

This source is identified with the Be star [MA93]1036.  Its longterm light
curve (Fig.\ 1) clearly shows the star undergoes repeated outbursts of
$\sim$0.05 mag.  The outburst (orbital) period is 127.3$\pm$1.0 days with
the epoch of maximum light at JD~2452195.2.  The periodogram and folded
light curve are plotted in Fig.\ 4.  We also searched for shorter periods
in the data, using both the entire data set and only the data from phases
between outbursts.  No clear short periods were identified. 

\subsection{ CXOU~J005750.3$-$720756 = SXP152 }

Macomb et al.\ (2003) found this source to be a 152-sec X-ray pulsar using
$Chandra$ data.  The position is coincident with the Be star [MA93]1038.
Its OGLE-III longterm light curve (Fig.\ 1) shows a gradual rise from
$I$=15.38 to 15.32 over four years, with considerable scatter in the last
two seasons.  Analysis of the flattened light curve shows no clear periods.
Some sections of the data have power near 70, 130, and 200 days, but these
may be artifacts due to the length of data train or spacing between
seasons.  We were unable to find any consistent period which fit all of
the data. 

\subsection{ RX~J0059.2$-$7138 = SXP2.76 }

The longterm light curve for this short-period pulsar is shown in Fig.\ 1.
Even before formal analysis one can detect repeated low-amplitude
outbursts.  The data were flattened in two segments (first three seasons
and final two seasons).  Both the individual segments and the full data
set give P=82.1$\pm$0.4 days, with the epoch of maximum at JD~2452189.4.
The outburst amplitude is $\sim$0.02 mag, with some variation from cycle
to cycle.  The folded light curve (Fig.\ 5) shows a rapid rise and slower
decline.  We have also searched for shorter periods using both the full
data set and data from out-of-outburst phases, but we found no significant
short periods that might be due to NRP. 

Based on Corbet's (1984) relation between pulse and orbital periods, this
source's orbital period is much longer than would be expected.  We suggest
that the pulsar may have been spun up by accretion of material from the Be
star's disk.

\subsection{ XTE~J0111.2$-$7317 = SXP31.0 }

This X-ray transient was identified by Coe et al.\ (2000) with a B0-2
emission-line star in the SMC.  They give $I$=15.31 on 1999 January 21,
which is in general agreement with the range of magnitudes ($I$=15.1-15.3)
observed in the OGLE-III survey.  Coe et al.\ found an irregular, small
emission nebula surrounding the star which they suggest might be a
supernova remnant.  Covino et al.\ (2001) obtained spectra and found this
star to have hydrogen emission lines, confiming its identification as a Be
star. 

Analysis of the OGLE-III photometry reveals a pronounced orbital period at
P=90.4$\pm$0.5 days, with an epoch of maximum at JD~2452808.9.  The
periodogram and PDM variance are plotted in Fig.\ 6, with both clearly
showing this orbital period.  However, the PDM variance gives the better
value since the light curve is very non-sinusoidal.  The folded light
curve has an extremely rapid climb to peak outburst and a more gradual
decay.  We have searched the inter-outburst data for any shorter periods,
but found none. 

\subsection{ RX~J0117.6-7330 = SXP22.1 }

Macomb et al.\ (1999) found this X-ray transient to be a 22-sec pulsar. It
is identified with the Be Star [MA93]1845.  Soria (1999) studied the
optical counterpart before the source was know to contain a pulsar.  He
gives the spectral type as B0.5IIIe with mean magnitudes of
$V$=14.9$\pm$0.01 and $I$=14.10$\pm$0.01.  We show the longterm OGLE-III
light curve in Fig.\ 1.  It is what we would call a ``moderate swooper",
with longterm variations from $I\sim13.9$ to 14.1 over several years. 

Analysis of the data from the first three seasons (the part which is most
easily flattened) reveals no clear periods that are not related to the
length of the data train or to seasonal gaps in the data.  The last two
seasons were more difficult to prewhiten, but these also did not show any
significant periods.  Finally, all of the flattened data were combined,
but again no photometric periods were found. 

\section{ Summary}

In summary, we have studied the OGLE-III light curves of eight Be/X-ray
pulsars in the SMC.  We have found orbital periods for four systems.  The
folded light curves all show asymmetrical outbursts with the rise time
being shorter than the decline.  One of the sources (SXP169) shows both
orbital outbursts and strong nonradial pulsations.  For the remaining
stars, no clear periodic behavior was found in the range 0.25-1000 days. 
The light curves of two stars near the position of SMC X-2 indicate that
the northern (brighter) object is the likely optical counterpart. 

In a recent paper on Be/X-ray binaries (Schmidtke \& Cowley 2006) we
presented some preliminary results about the photometric behavior of this
type of source.  The same general trends persist in the present
investigation.  We find that systems with strong orbital outbursts usually
do not exhibit nonradial pulsations of the Be star, and conversely the
sources with strong NRP rarely show orbital outbursts.  SXP169 is the
single exception to this.  We also find that systems with ``swooping''
light curves may show weak orbital outbursts, but don't have NRP. 

The strength of the orbital outbursts may be related to the eccentricity
of the orbit, if the pulsar penetrates the Be star's disk near periastron.
Perhaps the NRP sources are in fairly circular orbits with little
interaction between the stars, and hence orbital outbursts are not
present.  To investigate these ideas, spectroscopic studies are needed.
Finally, we note that Corbet's (1984) relation between pulse period and
orbital period is generally supported here, but it shows considerable
scatter.  Of note is SXP2.76 which shows an unexpectedly long orbital
period (P$_{orb}$=82.1 days) for its short X-ray pulse period, perhaps
indicating that the pulsar has been spun up. 

\acknowledgments

Support for the OGLE project was provided by the following grants: Polish 
MEiN grant 2P03D02124, NSF grant AST-0204908, and NASA grant NAG5-12212.

\newpage

\begin{figure}
\caption{ Longterm OGLE-III light curves of all the Be/X-ray pulsar
systems studied in this paper.  In the case of SMC X-2, for which the
identification is still uncertain, we have included the light curves for
both components.  The northern star is a ``swooper", a photometric
behavior which is found in many sources of this type. } 
\end{figure}

\begin{figure}
\caption{ (top) PDM variance (upper; right scale) and periodogram (lower;
left scale) for RX~J0052.9$-$7158 (SXP169).  (second) Light curve
folded on the orbital period of 67.6 days and binned in 20 phase bins.
(third) Periodogram of non-outburst data revealing the strong NRP at 0.755
days.  (bottom) Light curve of non-outburst data folded on P=0.755 days. }
\end{figure}

\begin{figure}
\caption{ Finding chart for SMC X-2 (SXP2.37) showing two stars which
are separated by $\sim2.5^{\prime\prime}$.  North is at the top; east is
on the left.  The figure is $\sim1.3^{\prime}$ on each side.  The light
curves of both components are shown in Fig.\ 1.  See discussion of the
optical identification in the text. } 
\end{figure}

\begin{figure}
\caption{ (top) Periodogram of RX~J0057.8$-$7202 (SXP280) showing the
strong period at 127.3 days.  The outbursting behavior is clearly visible
even without formal analysis (see Fig.\ 1).  (bottom) Light curve folded
on the orbital period of P=127.3 days.} 
\end{figure}

\begin{figure}
\caption{ (top) Periodogram of RX~J0059.2$-$7138 (SXP2.76) showing
power at P=82.1 days.  (bottom) Light curve folded on the orbital period
of P=82.1 days.
} 
\end{figure}

\begin{figure}
\caption{ (top) PDM variance (upper; right scale) and periodogram (lower;
left scale) for XTE~J0111.2$-$7317 (SXP31.0).  Both techniques show
power at P=90.4 days which appears to be the orbital period.  (bottom)
Light curve folded on the orbital period, showing prominent outbursts. } 
\end{figure}

\newpage

\begin{deluxetable}{llll}
\tablenum{1}
\tablecaption{Small Magellanic Cloud Be/Neutron-Star Systems Studied}
\tablehead{  
\colhead{System Name} & 
\colhead{SXP \#} &
\colhead{[MA93]\tablenotemark{a}} &
\colhead{2000 Position\tablenotemark{b}} 
}
\startdata
RX~J0052.9$-$7158      & SXP169  & 623   & 00:52:55.28 ~ $-$71:58:06.0 \nl
SMC X-2 (North)        & SXP2.37 & - - - & 00:54:33.47 ~ $-$73:41:01.1 \nl
SMC X-2 (South)        & SXP2.37 & - - - & 00:54:33.37 ~ $-$73:41:03.5 \nl
2E~0053.7$-$7227       & SXP34.1 & - - - & 00:55:28.46 ~ $-$72:10:56.5 \nl
RX~J0057.8$-$7202      & SXP280  & 1036  & 00:57:49.58 ~ $-$72:02:35.7 \nl
CXOU~J005750.3$-$720756 & SXP152 & 1038  & 00:57:50.38 ~ $-$72:07:55.9 \nl
RX~J0059.2$-$7138      & SXP2.76 & - - - & 00:59:12.74 ~ $-$71:38:44.8 \nl
XTE~J0111.2$-$7317     & SXP31.0 & - - - & 01:11:08.53 ~ $-$73:16:46.0 \nl
RX~J0117.6-7330        & SXP22.1 & 1845  & 01:17:40.18 ~ $-$73:30:50.5 \nl
\enddata

\tablenotetext{a}{From catalogue of SMC H$\alpha$ emission-line stars
by Meyssonnier \& Azzopardi (1993). } 

\tablenotetext{b}{2000 positions are from OGLE-III data.}

\end{deluxetable}

\newpage

\begin{deluxetable}{lccl}
\tablenum{2}
\tablecaption{Photometric Periods for 
SMC Be/Neutron-Star Systems}
\tablehead{  
\colhead{System Names} &
\colhead{P$_{orb}$} &
\colhead{NRP\tablenotemark{a}} &
\colhead{References} \\
\colhead{} &
\colhead{(days)} & 
\colhead{(days)} & 
\colhead{}
}
\startdata
RX~J0052.9$-$7158 (SXP169) & 67.6 & 0.755 & 
Galache et al.\tablenotemark{b} \nl
RX~J0057.8$-$7202 (SXP280) & 127.3 & - - \nl
RX~J0059.2$-$7138 (SXP2.76) & 82.1 & - - \nl
XTE~J0111.2$-$7317 (SXP31.0) & 90.4 & - -  & Coe et al.\tablenotemark{c} \nl

\enddata

\tablenotetext{a}{NRP = nonradial pulsations of the Be star. } 

\tablenotetext{b}{Galache et al.\ (2005) derived P$_{orb}$=68.6 days
from recurring X-ray outbursts. }

\tablenotetext{c}{Coe et al.\ (2000) found a small emission nebula 
surrounding this system. }

\end{deluxetable}

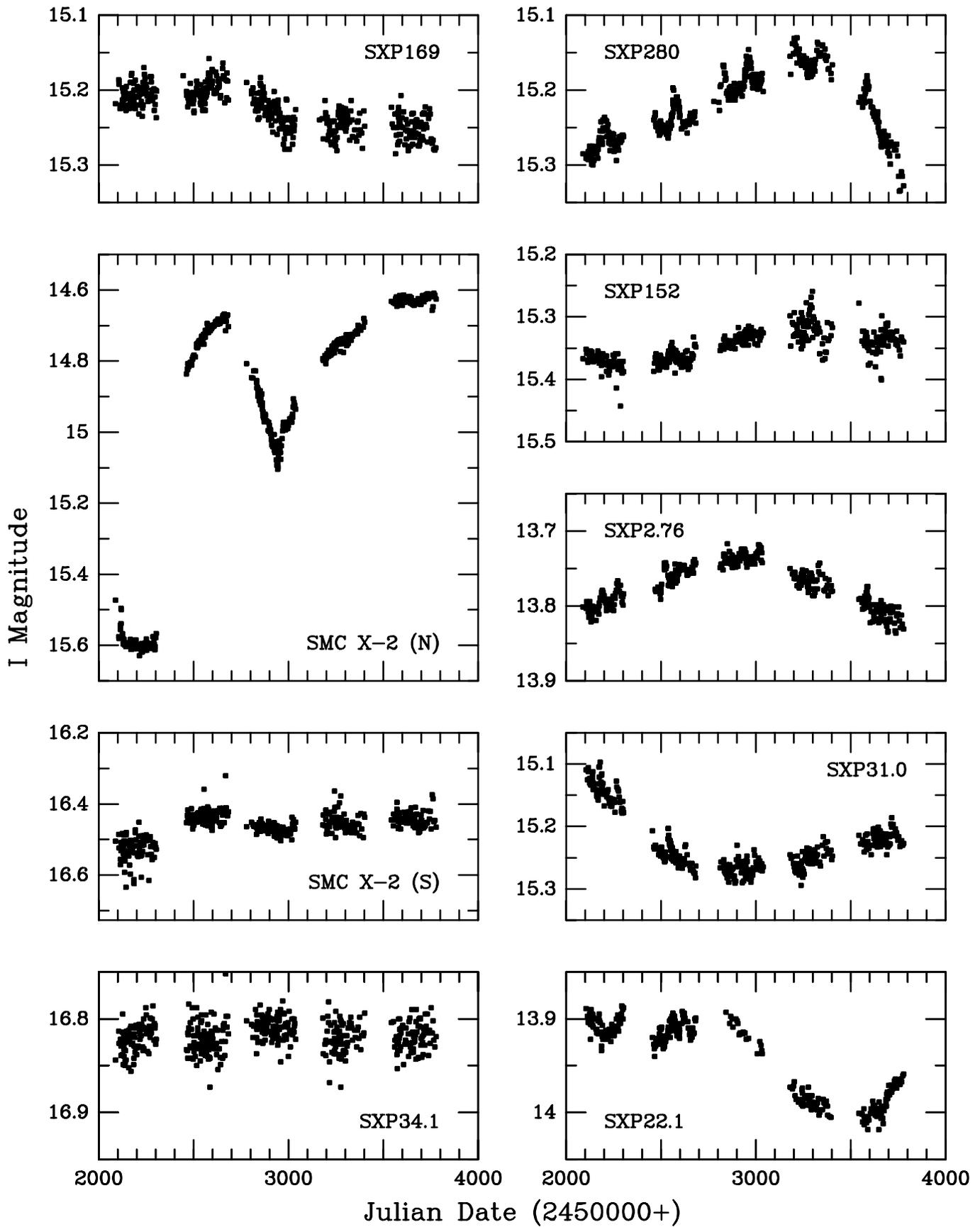
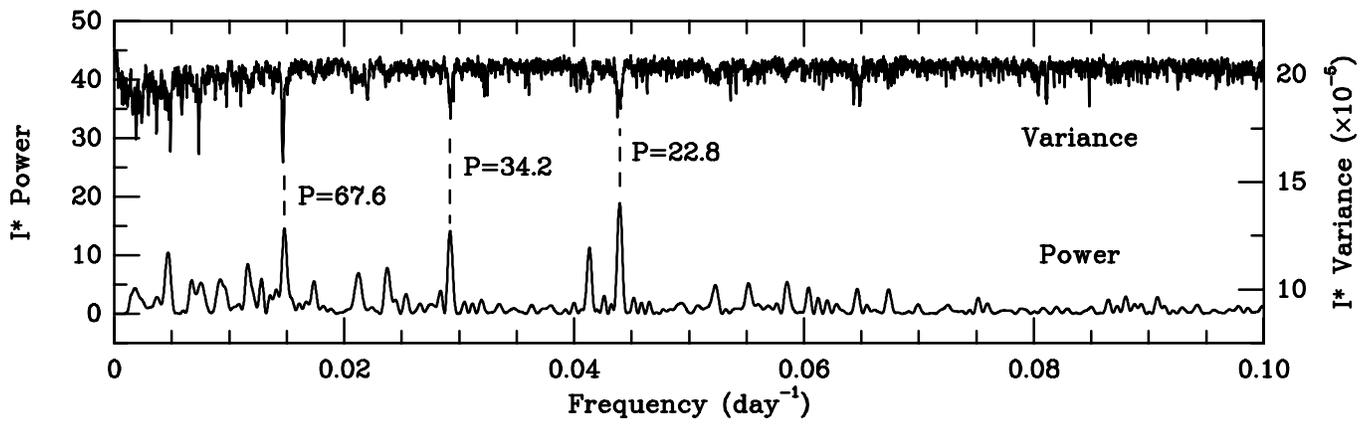
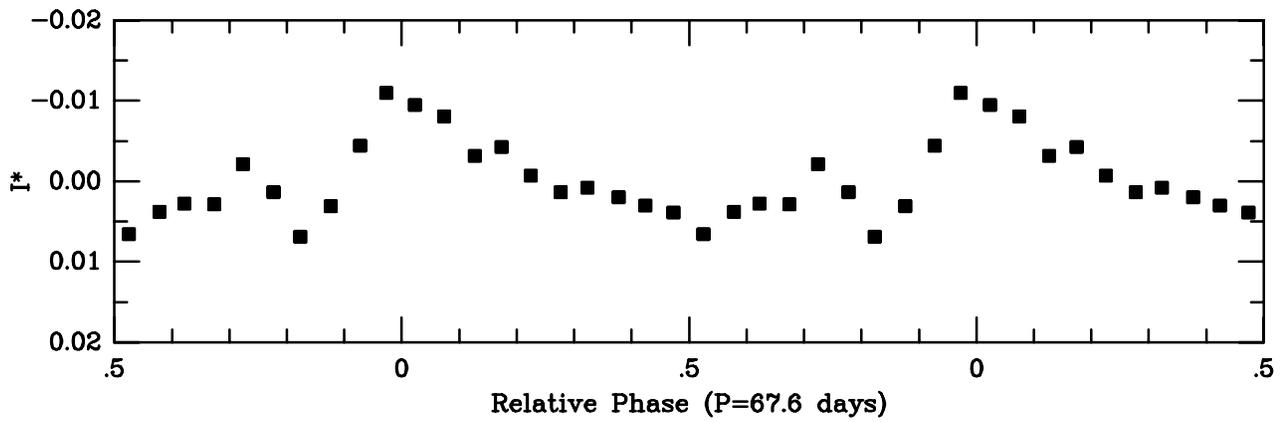
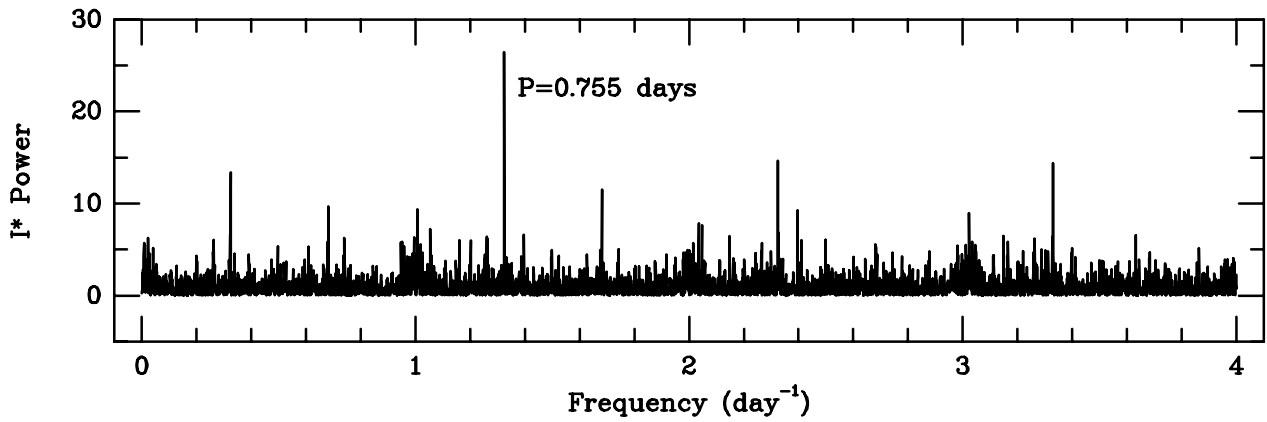
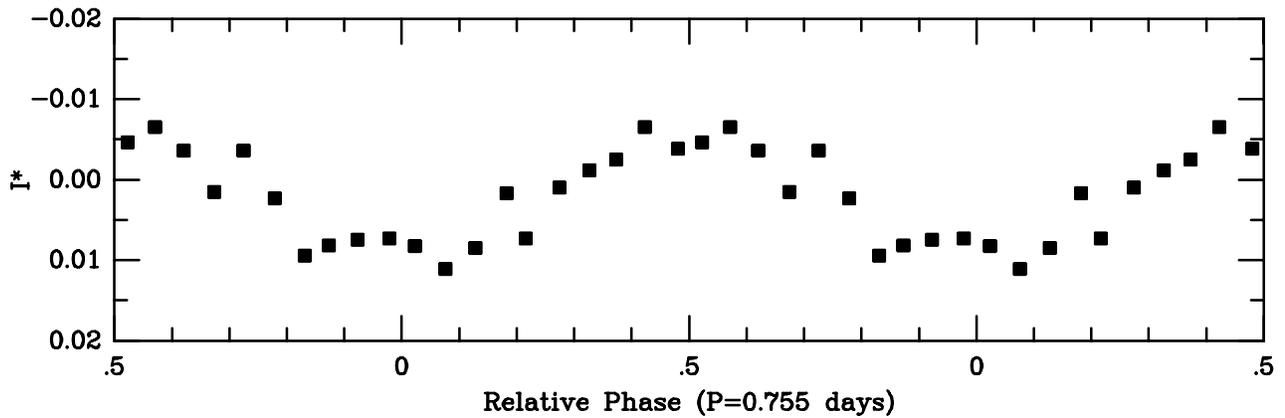
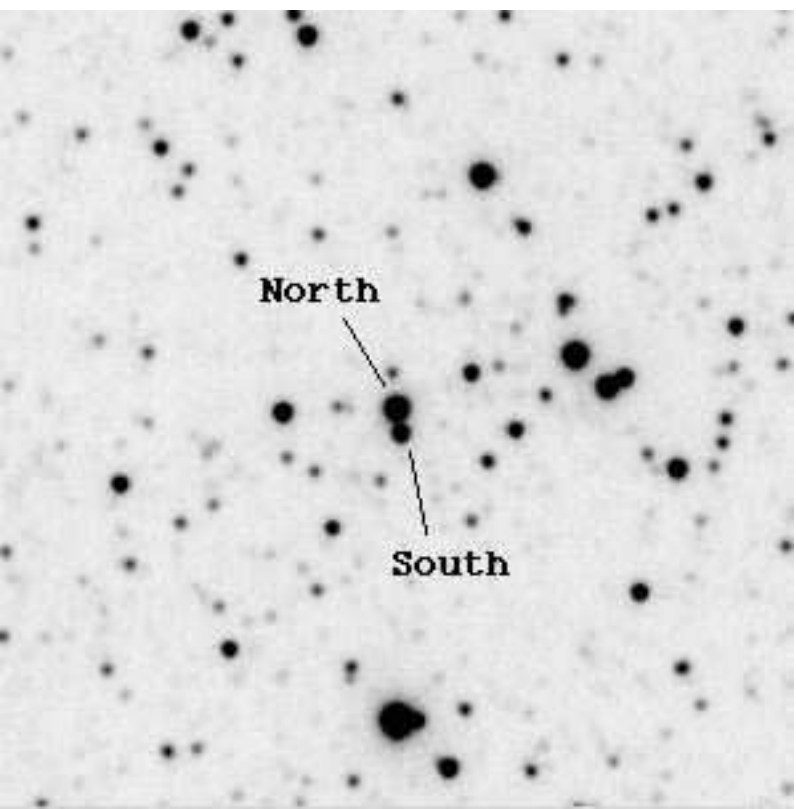
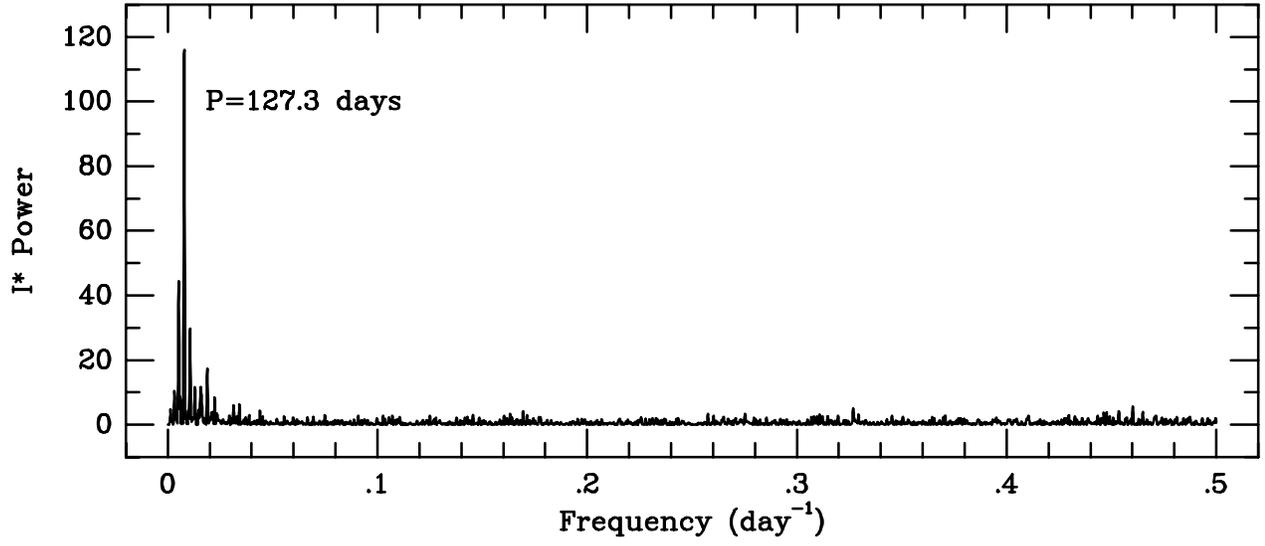
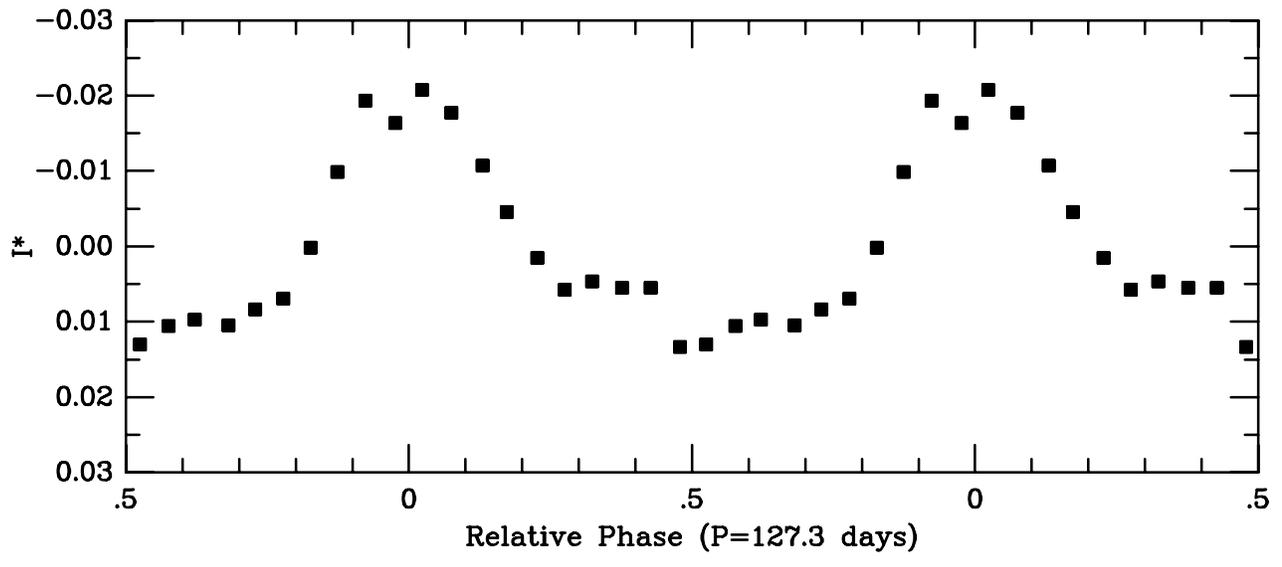
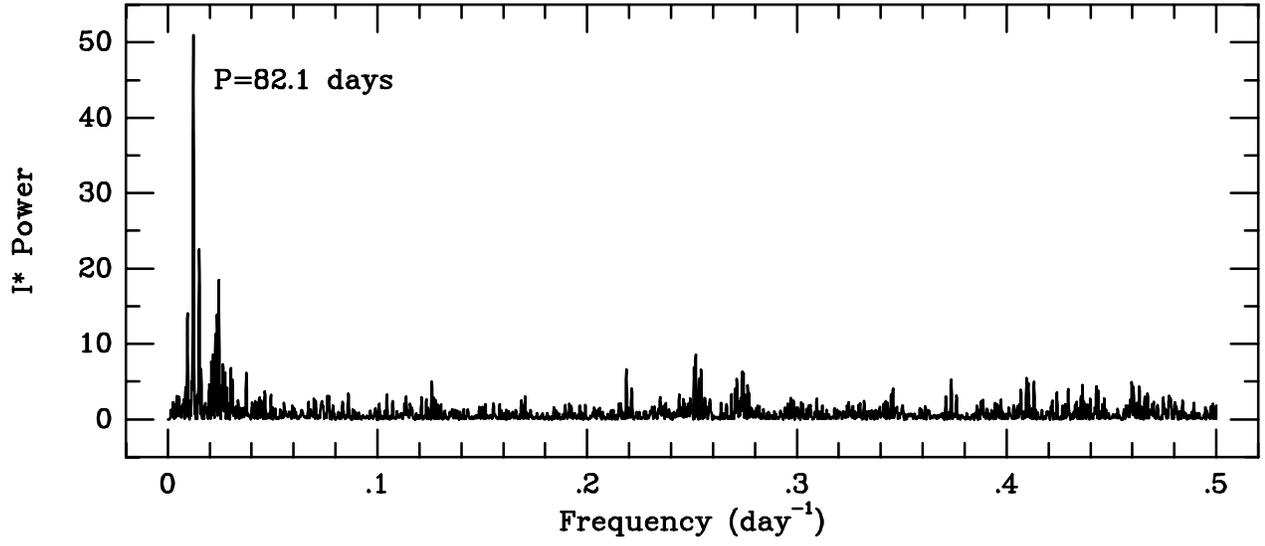
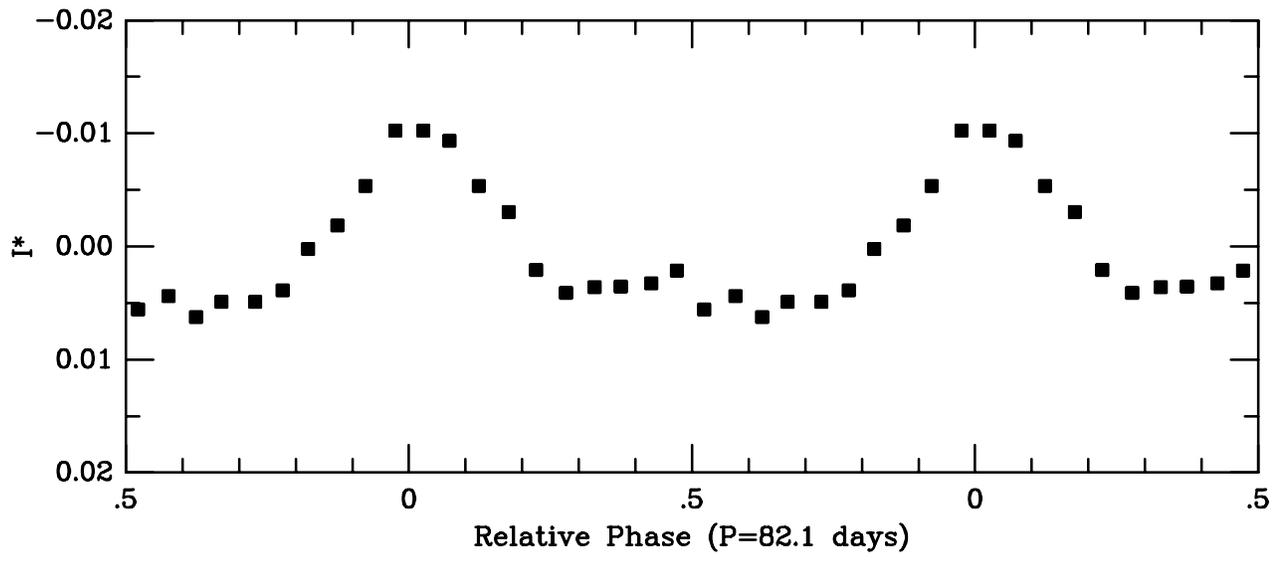
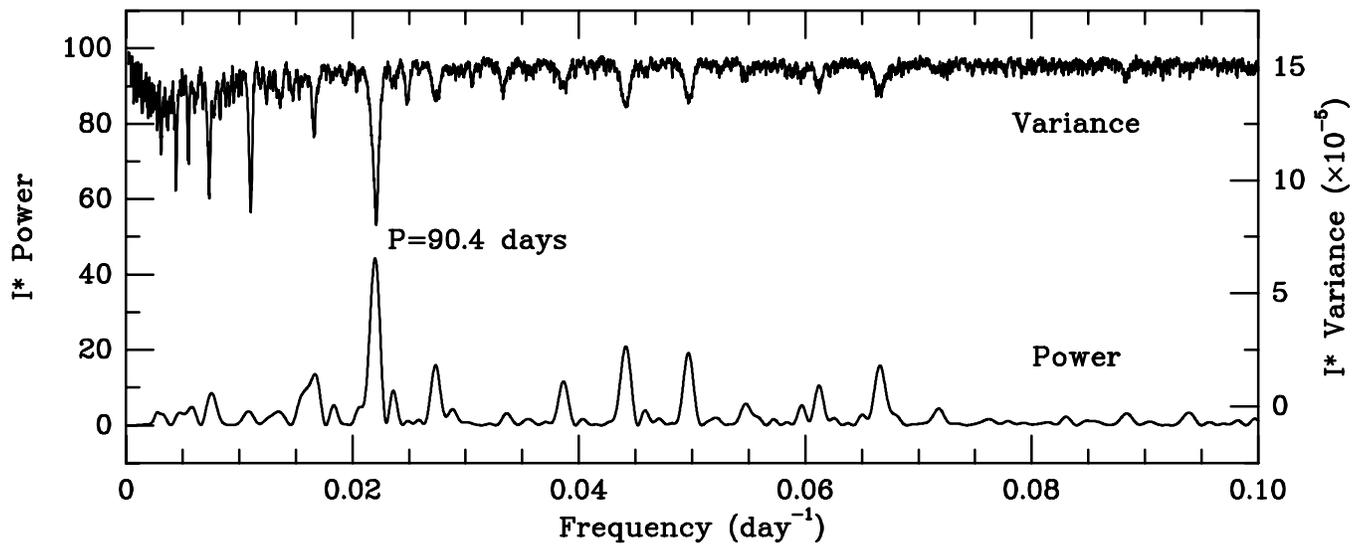
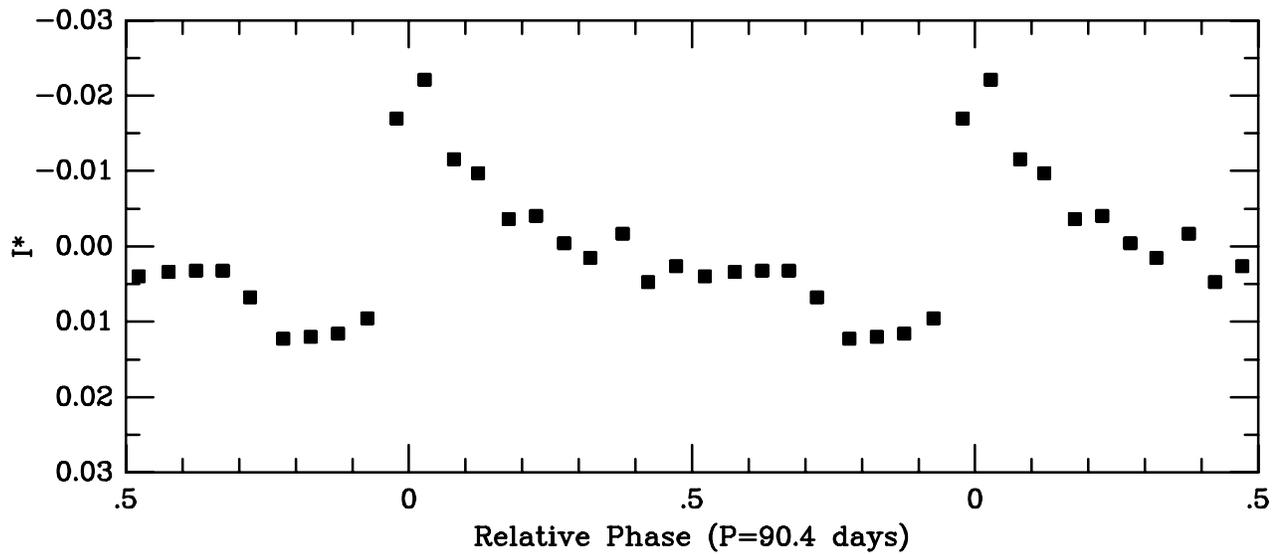
\end{document}